\documentclass[twocolumn,eqsecnum,aps]{revtex4}

\usepackage[dvips]{epsfig,color}
\usepackage{graphicx}
\usepackage{amsmath,amsfonts,amsbsy,amssymb}
\usepackage{tabularx}

\begin{document}

\title{Unification of  Laughlin and Moore-Read States in 
  SUSY Quantum Hall Effect}

\author{Kazuki Hasebe}
\affiliation{Department of General Education, Takuma National College of Technology,   Takuma-cho, Mitoyo-city, Kagawa 769-1192, Japan \\
Email: hasebe@dg.takuma-ct.ac.jp}

\begin{abstract}
Based on the recently proposed SUSY quantum Hall effect,  we show that   
   Laughlin  and  Moore-Read states are related by a hidden SUSY transformation.
Regarding the SUSY Laughlin wavefunction as a master wavefunction, Laughlin and Moore-Read states appear as two extreme limits of  component wavefunctions. 
Realizations of topological excitations on  Laughlin and  Moore-Read states
are also discussed in the SUSY formalism.
We  develop a streographically projected formulation of the SUSY quantum Hall effect.
With appropriate interpretation of Grassmann odd coordinates,  we illustrate 
 striking analogies between  SUSY quantum Hall effect and  superfluidity. 
\end{abstract}

\maketitle

  Quantum Hall effects (QHE) have been experimentally
 confirmed at  odd-denominator fillings and  even-denominator fillings 
 in monolayer and bilayer systems \cite{PRL481559,PRL591776,PRL9016801,PRL681383}. 
In monolayer QHE at 
 odd denominator filling, Laughlin wavefunction 
\cite{LaughlinPRL50} well describes the groundstate.
 For other filling QHE, several 
``deformed''  Laughlin wavefunctions have been proposed as  groundstates.
 Halperin wavefunction is a two-component extension of the Laughlin wavefunction and is introduced to formulate two-spin or bilayer QH liquids \cite{Halperin5675}.
 Especially,  Halperin 331 state is believed to describe the bilayer QH liquid at $\nu=1/2$. 
 Moore-Read state \cite{NPB360362,PRL663205} and  Haldane-Rezayi state \cite{PRL60956}  are proposed as  groundstate for even-denominator filling monolayer QH state.
The numerical calculation \cite{PRL801505}  and the recent experiment  \cite{cond-mat/0103144} suggest that  
 Moore-Read state is a probable groundstate at $\nu=5/2$.
 Moore-Read and Haldane-Rezayi states are constructed by multiplying the pfaffian  and the permanent factors to Laughlin and Halperin $331$ states, respectively.

Based on the composite boson \cite{PRL6282,PRB467765} and composite fermion 
\cite{PRL63199} picture, one may find  
 close analogies between QHE and superfluidity.
 Laughlin state is regarded as  ``condensation''
 of  composite bosons, analogous to 
 $^4He$ superfluidity. 
Haldane-Rezayi state describes condensation of $d$-wave pairing states of composite fermions.
Both Moore-Read  and  Halperin 331 states represent  
condensation  of  $p$-wave pairings of  composite fermions.
Besides, intriguingly, Moore-Read and Halperin 331 states  
 are simply related by a continuous rotation of the pairing vector, and their symmetries correspond to those of  the  superfluid $^3He-A_1$ and $^3He-A$ phases, respectively \cite{cond-mat/9503008}. 
 The symmetry and superfluid analogy arguments have discovered close relations between  such ``different''  QH liquids.

In this paper, we discuss a hidden SUSY relation between  Laughlin and  Moore-Read states. 
We use a set-up of the SUSY QHE \cite{hep-th/0411137,hep-th/0503162}, which was recently proposed as   a SUSY extension of the Haldane's spherical 
 QHE \cite{HaldanePRL51}.
The SUSY QHE contains non-anticommutative geometry as its mathematical background  
 \cite{NPB70994} and its low energy sector is described by a particular SUSY Chern-Simons field theory \cite{hep-th/0606007}.
Though the SUSY QHE may not   have apparent 
  relevance to  real QHE,  
 the SUSY QHE provides an interesting unified formulation of 
 Laughlin and Moore-Read states. Further, with appropriate interpretation of the Grassmann odd coordinates, like other QH liquids,  we show  the SUSY QH liquid may be understood as an exotic superfluid state.

\vspace{2mm}
{\it{SUSY Quantum Hall Effect}---}
The spherical SUSY QHE is constructed on a supersphere  in 
a supermonopole background \cite{hep-th/0411137}.
The supersphere $S^{2|2}=OSp(1|2)/U(1)$ is a supermanifold 
  whose coordinates satisfy the constraint 
$x_a^2+C_{\alpha\beta}\theta_{\alpha}\theta_{\beta}=R^2$, where $x_a (a=1,2,3)$ and $\theta_{\alpha} (\alpha=1,2)$ are Grassmann even and odd coordinates, respectively,
 and $C_{\alpha\beta}$ is a charge conjugation antisymmetric matrix 
with $C_{12}=1$.
The one-particle Hamiltonian  is given by 
$H=\frac{1}{2MR^2}(\Lambda_a^2+C_{\alpha\beta}\Lambda_{\alpha}\Lambda_{\beta}),$
where $\Lambda_a=-i\epsilon_{abc}x_b D_c+\frac{1}{2}\theta_{\alpha}(\sigma_a)_{\alpha\beta}D_{\beta}$ and $\Lambda_{\alpha}=\frac{1}{2}(C\sigma_a)_{\alpha\beta}x_aD_{\beta}-\frac{1}{2}\theta_{\beta}(\sigma_a)_{\beta\alpha}D_a$ denote the covariant angular momenta. The covariant derivatives,  
$D_a=\partial_a-iA_a$ and  $D_{\alpha}=\partial_{\alpha}-iA_{\alpha}$,  
satisfy the algebras,   
$[D_a,D_b]=i\frac{I}{2R^3}\epsilon_{abc}x_c(1+\frac{3}{2R^2}{C_{\alpha\beta}\theta_{\alpha}\theta_{\beta}})$, $[D_a,D_{\alpha}]=\frac{I}{2R^3}( \delta_{ab}-\frac{3}{R^2}{x_a x_b}) (\sigma_b C)_{\alpha\beta}\theta_{\beta}$ and 
$\{D_{\alpha},D_{\beta}\}=\frac{I}{R^3}x_a(\sigma_a C)_{\alpha\beta}(1+\frac{3}{2R^2}C_{\gamma\delta}\theta_{\gamma}\theta_{\delta})$.
 The  energy eigenvalues are given by 
$E_{n}=\frac{1}{2MR^2}[n(n+\frac{1}{2})+(n+\frac{1}{4})I]$,
where $I/2$ indicates the half-integer charge of the supermonopole  and $n$ is 
an integer to specify Landau level. 
The degeneracy in $n$-th  Landau level is given by $D_{n}
=4n+2I+1$. Especially,  the degeneracy in the lowest Landau level (LLL)
 is  $2I+1$, and the degenerate wavefunctions are given by 
supermonopole harmonics which are constructed by the  
products of the components of the  super Hopf spinor 
$\psi=(u,v,\eta)^t=(\sqrt{\frac{1+x_3}{2}}(1-\frac{1}{4(1+x_3)}
C_{\alpha\beta}\theta_{\alpha}\theta_{\beta}),\frac{x_1+ix_2}{\sqrt{2(1+x_3)}}(1+\frac{1}{4(1+x_3)}C_{\alpha\beta}\theta_{\alpha}\theta_{\beta}),\sqrt{\frac{1+x_3}{2}}\theta_1+
\frac{x_1+ix_2}{\sqrt{2(1+x_3)}}\theta_2)^t$
 as  
\begin{equation}
u^{I-p}v^p,~~~u^{I-q-1}v^{q}\eta,\nonumber
\end{equation}
with $0\le p \le I,~~0\le q\le I-1$.
Their eigenvalues of $L_3$ are given by $\frac{I}{2}-p$ and $\frac{I}{2}-q-\frac{1}{2}$, respectively.
The spherical SUSY Laughlin wavefunction is derived as   
\begin{equation}
\Psi
=\prod_{i<j}^N(u_iv_j-v_iu_j-\eta_i\eta_j)^m, 
\nonumber
\label{susyllinsphere}
\end{equation}
and it is  invariant under the $OSp(1|2)$ transformation 
 generated by 
$L_a=\psi^t l_a\frac{\partial}{\partial\psi}$ and $L_{\alpha}=\psi^t l_{\alpha}\frac{\partial}{\partial\psi}$. Here, $l_a$ and $l_{\alpha}$ are given by 
$l_a=\frac{1}{2}
\begin{pmatrix}
\sigma_a & 0\\
0 & 0
\end{pmatrix},~~
l_{\alpha}=\frac{1}{2}
\begin{pmatrix}
0 & \tau_{\alpha}\\
-(C\tau_{\alpha})^t & 0
\end{pmatrix}$,
where  $\tau_1=(1,0)^t$ and $\tau_2=(0,1)^t$. 
Since the original Laughlin wavefunction takes the 
form \cite{HaldanePRL51}
\begin{equation}
\Phi=\prod_{i<j}(u_{i}v_{j}-v_i u_j)^m,\nonumber
\end{equation} 
 the only difference between $\Psi$ and $\Phi$ is 
 the  $\eta\eta$ term  \cite{comment1}.
As the SUSY system generally contains both bosonic and fermionic states, the SUSY QHE at $\nu={1}/{m}$ is doubly degenerate compared to the original ``bosonic'' QHE at $\nu={1}/{m}$ \cite{hep-th/0411137, hep-th/0503162}.

Here, we introduce interesting correspondences between the SUSY QHE and 
 a projected original QHE which we call the ``vector'' QHE.
We focus on  the original Haldane's system subject to {\it{even}} Landau levels 
 in  the background of  Dirac monopole with {\it{integer}} charge.
In such  projected Haldane's system, the energy eigenvalues are given by $\epsilon_{2n}=4\times \frac{1}{2MR^2}(n(n+\frac{1}{2})+I(n+\frac{1}{4}))$, which are exactly equivalent to the energy spectra for the above SUSY Landau problem up to the proportional factor 4. Besides, the degeneracy in $2n$-th Landau level is given by $d_{2n}=
4n+2I+1$, which is again equal to the degeneracy of  $n$-th SUSY Landau level.
Thus, the original Landau problem subject to 
 even Landau levels with  integer monopole charge  has  $\it{quantitative}$ correspondence   
 to the SUSY Landau problem.
The degeneracy of the present LLL  is 
$d_0=2I+1$, and the degenerate eigenstates are given by the vector monopole harmonics 
\begin{equation}
 U^{I-p}V^{p},~~
U^{I-q-1}V^{q} W, \nonumber
\end{equation}
where  $0\le p\le I,~~0\le q\le I-1$, and  $(U,W,V)=(u^2,\sqrt{2}uv,v^2)$ denotes the 
SO(3) Hopf vector. The eigenvalues of $L_3$ for the above vector monopole harmonics 
 are, respectively, given by 
 $I-2p$ and $I-2q-1$. One may find   apparent analogies between the vector monopole harmonics and the super monopole harmonics  on the correspondence  between the Hopf vector $(U,V,W)$ 
 and  the SUSY Hopf spinor $(u,v,\eta)$.  
Since  $SO(3)$ singlet is made as the symmetric combination of two vectors,
  the ``vector''  Laughlin wavefunction  is constructed as  
\begin{equation}
{\Phi}_v=\prod_{i<j}( U_iV_j+V_iU_j-W_iW_j )^m 
=\!\!\prod_{i<j}(u_iv_j-u_i v_j)^{2m}.\nonumber
\end{equation}
Again, there is a  formal resemblance between $\Psi$ and $\Phi_v$
 by the correspondence between $(U,V,W)$ and $(u,v,\eta)$.
The vector Laughlin wavefunction is simply equal to the original Laughlin wavefunction at the even 
 denominator filling   $\nu={1}/{2m}$, 
   which  
 contains double states compared to the original system at $\nu=1/m$.
 This may reminds  the doubly degenerate feature of   
 SUSY system.
 Thus,  at the many-body wavefunction  level, the vector QHE possesses  
 several qualitative  analogies to the SUSY QHE.

\vspace{2mm}
{\it{Expansion of SUSY Laughlin State.---}}
 We expand 
the SUSY Laughlin wavefunction  in terms of the Grassmann odd quantities  
 as in the superfield formalism.
With the composite quantity 
\begin{equation}
Q=\sum_{i<j}\frac{\eta_i\eta_j}{u_{i}v_{j}-v_i u_j},\nonumber
\end{equation}
which we call the pairing operator, 
the SUSY Laughlin wavefunction is simply rewritten as  
\begin{equation}
\Psi
=e^{-m Q }\cdot\Phi=\sum_{k=0}^{N/2}(-m)^k \Psi_k.\nonumber
\label{BCSSUSYLlin}
\end{equation}
In the last equation, we  expanded the exponential in terms of $Q$, and  
 the component  wavefunctions   
$\Psi_k$  $(k=1,2,\cdots, N/2)$ are given by 
\begin{equation}
\Psi^{}_k=\frac{1}{k!}Q^k\cdot \Phi.
\label{basicpsikphi1}\nonumber
\end{equation}
Explicitly, they are 
\begin{subequations}
\begin{align}
&\Psi^{}_0 =
\Phi^{},~~~~\nonumber\\
&\Psi_1^{}
=\sum_{i<j}\eta_i\eta_j \frac{1}{{\phi}_{ij}}\cdot \Phi,\nonumber\\
& \Psi_2^{}=
\sum_{i<j<k<l} \!\!\!\!\!\!\eta_i\eta_j\eta_k\eta_l 
  \frac{1}{{\phi}_{ijkl}} \cdot \Phi^{},\nonumber\\
&\vdots\nonumber\\
&\Psi_{{N}/{2}}^{}= 
   \eta_1\eta_2\cdots \eta_N \cdot
{Pf}(\frac{1}{\phi_{ij}})~\Phi^{},\nonumber
\end{align}
\end{subequations}
where 
$ \frac{1}{\phi_{ij}} \equiv \frac{1}{u_i v_j-v_i u_j},
 \frac{1}{\phi_{ijkl}} \equiv
\frac{1}{\phi_{ij}\phi_{kl}}
 -
 \frac{1}{\phi_{ik}\phi_{jl}}
+\frac{1}{\phi_{il}\phi_{jk}}.$
In the  expression of  $\Psi_{N/2}$,   we used the formula 
$\eta_{i_1}\eta_{i_2}\cdots \eta_{i_N}=\epsilon_{i_1i_2\cdots i_N}\eta_1\eta_2\cdots \eta_N$. It is noted that the original Laughlin wavefunction appears as the 1st component
 wavefunction $\Psi_0$, and, remarkably, 
 Moore-Read state 
  appears as the last component wavefunction $\Psi_{N/2}$.
This expansion suggests that Laughlin and Moore-Read states are related
 by a  SUSY transformation.
To clarify the direct SUSY relation between Laughlin and  Moore-Read states, 
 it is important to explore the properties of the pairing operator.
While $Q$ is a $SU(2)$-singlet quantity, it is not invariant under the 
Grassmann odd transformation generated by $L_{\alpha}$.
However,  as is easily checked,  $Q$   returns to itself under the two successive 
operations of $L_{\alpha}$,
\begin{equation}
 C_{\alpha\beta}L_{\alpha}L_{\beta} Q=\frac{1}{2}Q-\frac{1}{4}N(N-1).\nonumber
\end{equation}
Besides, since the SUSY Laughlin wavefunction $\Psi=e^{-mQ}\Phi$ is invariant under the $OSp(1|2)$ transformation, 
    $\Phi$ and $Q$ satisfy the relation 
$C_{\alpha\beta}L_{\alpha}L_{\beta}\ln\Phi=mC_{\alpha\beta}L_{\alpha}L_{\beta}Q.$
These relations  suggest that $Q$ is generated by the two successive SUSY  transformations from $\Phi$, 
\begin{equation}
Q= \frac{2}{m}C_{\alpha\beta}L_{\alpha}L_{\beta}\ln\Phi +\frac{N(N-1)}{2}.  \nonumber
\end{equation}
With this relation,  every component wavefunction can be  
represented by  the SUSY transformation of  the original Laughlin wavefunction.
For instance, the 2nd component wavefunction is expressed as  
$\Psi_1=Q\Phi= \frac{2}{m}C_{\alpha\beta}L_{\alpha}L_{\beta}\Phi
-\frac{2}{m}C_{\alpha\beta}(L_{\alpha}\ln\Phi)(L_{\beta}\ln\Phi)  \Phi 
+\frac{N(N-1)}{2}\Phi$. Repeating such  SUSY transformation $N/2$ times, 
 Moore-Read state  $\Psi_{N/2}$  is finally constructed from  the Laughlin wavefunction $\Phi$ as  
\begin{align}
&\Psi_{N/2}  = \frac{1}{(N/2)!} Q^{\frac{N}{2}} \Phi \nonumber\\
&~~~~~~=
\frac{1}{(N/2)!} \biggl(\frac{2}{m}\biggr)^{\frac{N}{2}}(C_{\alpha\beta}L_{\alpha}L_{\beta})^{\frac{N}{2}} \Phi +(\text{non-linear terms}).\nonumber
\end{align}
Thus, Laughlin and Moore-Read states are related by the nonlinear SUSY transformation generated by $L_{\alpha}$. 

\vspace{2mm}
{\it{Topological Excitations}---}
The quasi-hole excitation at the point $(\Omega_a,\Omega_{\alpha})=(2\chi^{\ddagger}l_a\chi,2\chi^{\ddagger}l_{\alpha}\chi)$ with $\chi=(a,b,\xi)^t$  and $\chi^{\ddagger}=(a^*,b^*,-\xi^*)$ 
  on the supersphere 
 is generated by the operator 
$\mathcal{A}^{\ddagger}
=\prod_i \mathcal{A}_i^{\ddagger},$
where  $\mathcal{A}_i^{\ddagger}= av_i-bu_i-\xi\eta_i 
$  \cite{hep-th/0411137}. It is apparent, without  $\xi\eta$ term, $A_{i}^{\ddagger}$ is reduced to 
$A_i^{\dagger}=av_i-bu_i$, which is the quasi-hole operator on the original Laughlin state. 
The fundamental excitation on the Moore-Read state is called the halberon, which carries half of the  electric charge of a naively expected quasi-particle excitation  \cite{PRL663205}. 
The halberon pair wavefunction is given by  $Pf\biggl(\frac{A_i^{\dagger}{A'_j}^{\dagger}+A_j^{\dagger}{A'_i}^{\dagger}} {\phi_{ij}} \biggr)\cdot\Phi$.
In the SUSY formalism, the halberon pair operator is constructed as  
$\mathcal{A}_{h.b.}^{\ddagger}= \prod_{i<j}\exp\biggl(m\frac{\eta_i\eta_j}{\psi_{ij}}(1-\mathcal{A}_i^{\ddagger}\mathcal{A}_j'^{\ddagger}- \mathcal{A}_j^{\ddagger} \mathcal{A}_i'^{\ddagger} ) \biggr),$ where $\psi_{ij}=u_iv_j-v_iu_j-\eta_i\eta_j$. Indeed, this operator  acts the SUSY Laughlin wavefunction as 
\begin{equation}
\mathcal{A}_{h.b.}^{\ddagger}\Psi=
\prod_{i<j}\exp\biggl(-m \frac{\eta_i\eta_j}{\phi_{ij}}{(A_i^{\dagger}A_j'^{\dagger}+ A_j^{\dagger}A_i'^{\dagger}) }\biggr)\cdot \Phi,\nonumber
\end{equation}
and $\mathcal{A}_{h.b.}^{\ddagger}\Psi$ yields the halberon pair wavefunction as the last term of its expansion.

\vspace{2mm}
{\it{The Stereographic Projection}}---
By the stereographic projection from the supersphere  to the superplane, 
we construct a planar SUSY QHE.  
The stereographic super-coordinates $(z,\theta)$ are defined as   
$z\equiv \frac{v}{u}=
\frac{x_1+ix_2}{R+x_3}
(R+\frac{1}{2(R+x_3)}C_{\alpha\beta}\theta_{\alpha}\theta_{\beta})$ and $\theta\equiv \frac{\eta}{u}=
\theta_1+\frac{1}{R}z\theta_2.$
$z$ and $\theta$ respectively indicate bosonic and fermionic complex coordinates on the superplane $R^{2|2}$.
With use of the stereographic super-coordinates, 
the super Hopf spinor   is rewritten as  
$\psi=(u,v,\eta)^t=\frac{1}{\sqrt{R^2+zz^*}}(R,z,\theta)^t(1-\frac{1}{2(R^2+zz^*)} \theta\theta^*)$, and in the thermodynamic limit $R, I\rightarrow \infty$ with $B=\frac{I}{2R^2}$ fixed, 
the supermonopole harmonics are 
 reduced to 
\begin{align}
&u^{I-p}v^{p}\rightarrow  z^{p}  \exp({-B(zz^{*}+\theta\theta^*)},\nonumber\\
&u^{I-q-1}v^{q}\eta\rightarrow z^{q}\theta\exp({-B(zz^{*}+\theta\theta^*)}.\nonumber
\end{align}
 These expressions  are consistent with the previously derived LLL basis 
  of the planar SUSY Landau problem \cite{hep-th/0503162}. 
Similarly, the planar SUSY Laughlin wavefunction is derived as 
\begin{equation}
\Psi=\prod_{i<j}(z_i-z_j+\theta_i\theta_j)^m\cdot\exp\biggl(-\frac{1}{\ell_B^2}\sum_i(z_iz_i^*+\theta_i\theta_i^*)\biggr),
\label{stereograLlin}\nonumber
\end{equation}
where  $\ell_B=R\sqrt{\frac{2}{mI}}$.  Some comments are added here.
While  the spherical SUSY Laughlin wavefunction is expressed
 as the bilinear form about Grassmann even and  odd quantities,
  the planar SUSY Laughlin wavefunction is linear for the bosonic coordinate $z$ 
and  bilinear for the fermionic coordinate $\theta$ (up to the exponential term).
As we shall see below, this discrepancy generates magnetic translation asymmetry about 
the bosonic and the fermionic directions in the planar case. 
The  $OSp(1|2)$ generators $L_a$ and $L_{\alpha}$    
correspond  to the magnetic translation generators 
$(X,Y,\Theta_1,\Theta_2)$ and the perpendicular angular momentum $L_{\perp}$ in the superplane case.
Physically, $(X,Y,\Theta_1,\Theta_2)$ represent the center of mass coordinates on the superplane, and  are defined as 
$(X,Y,\Theta_1,\Theta_2)=(x-i\ell_B^2D_y,y+i\ell_B^2D_x, \theta_1+\ell_B^2D_{\theta_2}, \theta_2+\ell_B^2 D_{\theta_1})$,  where the covariant derivatives  
$D_x={\partial_x}-iA_x, D_y={\partial_y}-iA_y,
 D_{\theta_1}=\partial_{\theta_1}-iA_{\theta_1}$ and  
$D_{\theta_2}=\partial_{\theta_2}-iA_{\theta_2}$  satisfy the relation
 $[D_x,D_y]=i\{D_{\theta_1},D_{\theta_2}\}=-iB$.
 $L_{\perp}$ is given by  
$L_{\perp} =L_{\perp}^B+L_{\perp}^F,$
where 
$L_{\perp}^B= -i\epsilon_{ij}x_i\frac{\partial}{\partial x_j}$ and $L_{\perp}^F=\frac{1}{2}\theta_{\alpha} (\sigma_3)_{\alpha\beta}\frac{\partial}{\partial \theta_{\beta}}$.
$L_{\perp}^B$ and $L_{\perp}^F$ represent the  {\it{orbital}} and $\it{spin}$ angular momentum operators, respectively \cite{hep-th/0503162}. 
In the symmetric gauge, these generators  are represented as   
$X= \frac{1}{2}(z+z^*+\ell_B^2\frac{\partial}{\partial z}-\ell_B^2\frac{\partial}{\partial z^*}),
Y=i\frac{1}{2}(-z+z^*+\ell_B^2\frac{\partial}{\partial z}+\ell_B^2\frac{\partial}{\partial z^*}),
\Theta_1=\frac{1}{\sqrt{2}}(\theta+\ell_B^2\frac{\partial}{\partial\theta^*}),
\Theta_2=\frac{1}{\sqrt{2}}(\theta^*+\ell_B^2\frac{\partial}{\partial\theta})$ 
 and $L_{\perp}=  z\frac{\partial}{\partial z}-z^*\frac{\partial}{\partial z^*}+\frac{1}{2}
(\theta\frac{\partial}{\partial \theta}-\theta^*\frac{\partial}{\partial \theta^*})$.
With these expressions, it is straightforward to prove  that 
the planar SUSY Laughlin wavefunction is  invariant only for the translations generated by $X-iY$, $(X-iY)\Psi=0$, and $L_{\perp}$, $L_{\perp}\Psi=  m\frac{N(N-1)}{2}\Psi.$ The stereographic projection breaks  the  translation symmetries 
 generated by $X+iY$ and all  fermionic generators $\Theta_1$ and $\Theta_2$ \cite{superwave}. 
The planar SUSY Laughlin wavefunction is zero energy state of the hard-core Hamiltonian 
\begin{equation}
H_{HC}=\sum_{l<m}V_{{l}} P_{{l}},\nonumber
\end{equation}
where  $l$ is the integer which indicates  the  eigenvalue of the 
dimensionless super-radius operator $
\frac{1}{2\ell_B^2}( X_i^2+C_{\alpha\beta}\Theta_{\alpha}\Theta_{\beta})$
 between arbitrary two-particles.

\begin{figure}[tbph]
\includegraphics*[width=90mm]{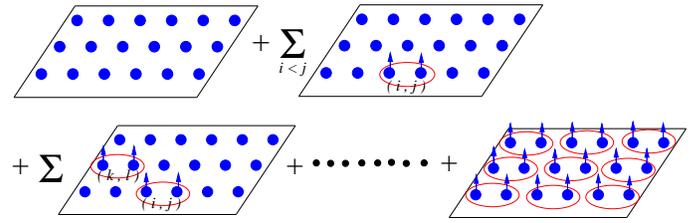}
\caption{The graphical representation of the expansion of the SUSY Laughlin wavefunction.}
\label{ExpandSUSYLlin}
\vspace{-3mm}
\end{figure}

 Replacing the  pairing operator with its  planar form
\begin{equation}
Q=-\sum_{i<j}\frac{\theta_i\theta_j}{z_{i}-z_{j}},\nonumber
\end{equation}
 the planar SUSY Laughlin wavefunction has same expansion  as of  
 the spherical SUSY Laughlin wavefunction. 
The pairing operator  and the angular momentum operators satisfy the relations,   
\begin{equation}
[L_{\perp}^B,Q]=-Q,~~[L_{\perp}^F,Q]=Q,~~[L_{\perp},Q]=0. \nonumber
\end{equation}
The physical meaning of  $Q$ is clear in this planar form. 
$\theta$ is the eigenstate of the spin operator $L_{\perp}^F$ with the eigenvalue $1/2$, and may be interpreted as the 1/2 spin-up state. (Similarly, $\theta^*$ may corresponds to the 1/2 spin-down state.)
 Meanwhile, $1/z$ represents a $p$-wave boundstate, since 
  $1/z$ carries  the orbital angular momentum  $-1$,  and 
 is the  eigenstate of the two-dimensional Schr$\ddot{\text{o}}$edinger equation of the delta-function type attractive potential.
Then, when $Q$ operates the Laughlin state, the numerator $\theta_i\theta_j$  attaches  
  $1/2$ up-spins to two spin-less fermions $i$ and $j$, and the denominator $\frac{1}{z_i-z_j}$ acts 
 to form a $p$-wave pairing state of such two particles.
Since $\Psi_1$ is constructed by multiplying $Q$ to $\Phi$  once, 
$\Psi_1$ contains one $p$-wave pairing state with polarized spins  on Laughlin state. Similarly, in $\Psi_2$, $Q$ is multiplied to $\Phi$ twice, and   two $p$-wave pairing states with polarized spins are formed  on the Laughlin state.
 Repeating this procedure, we finally obtain 
the last component wavefuncton $\Psi_{N/2}$, in which 
all  particles form $p$-wave pairing states with polarized spins  [Fig.\ref{ExpandSUSYLlin}]. This $p$-wave pairing superfluid state is indeed the  Moore-Read state, and $\Psi_{N/2}$ even catches  the polarized spin structure of Moore-Read state.   
The  $k$-th component wavefunction $\Psi_k$ 
is an eigenstate of both orbital angular momentum $L_{\perp}^B$ and spin angular momentum $L_{\perp}^F$ as 
\begin{align}
&L_{\perp}^B\Psi_k=(m\frac{N(N-1)}{2}-k)\Psi_k,\nonumber\\
&L_{\perp}^F\Psi_k= k \Psi_k.\nonumber
\end{align}
These relations  support the above physical interpretation of 
the component wavefunctions.
Since the original Laughlin state and the $k$ $p$-wave pairing states   have  the orbital angular momenta, $m\frac{N(N-1)}{2}$ and  $k\times(-1)$, respectively, 
 the total orbital angular momentum of $\Psi_k$ is given by $m\frac{N(N-1)}{2}-k$.
Similarly, since there are $k$ pairs of polarized spins  on the Laughlin state,
 the total spin angular momentum of $\Psi_k$ is $\frac{1}{2}\times {2}k$. Though  the component wavefunctions carry different orbital and spin angular momenta, each of them carries the identical 
angular momentum, $m\frac{N(N-1)}{2}$, in total.

\vspace{2mm}
{\it{Analogy to Superfluidity.---}}
The above expansion of the SUSY Laughlin state suggests close  analogies between  the superfluidity and the SUSY QHE. The BCS wavefunction has the form,  
$|\text{BCS}\!\!>=\prod_{k}(1+ g_k c^{\dagger}_{k\uparrow}c^{\dagger}_{-k\downarrow})|0\!\!>$, where $g_k$ is the coherence factor 
 and $c^{\dagger}_{k\sigma}$ indicates the creation operator for electron with momentum $k$ and spin $\sigma$, ${c_{k\sigma}^{\dagger}}^2=0,~\{c_{k\sigma},c^{\dagger}_{k'\sigma'}\}=\delta_{kk'}\delta_{\sigma\sigma'}$.  BCS state can be  expressed as the superposition of the number states of Cooper pairs \cite{ScriefferBook1964}, 
$|\text{BCS}\!\!> =
e^{\sum_k g_k c^{\dagger}_{k\uparrow}c^{\dagger}_{-k\downarrow}}|0\!\!>=|0\!\!>+\sum_k g_k c^{\dagger}_{k\uparrow}c^{\dagger}_{-k\downarrow}|0\!\!>+\cdots+
  \prod_k g_k c_{k\uparrow}^{\dagger}c_{-k\downarrow}^{\dagger}|0\!\!>.$
Comparing the expansion formulas of  BCS state and  SUSY Laughlin state, 
one may find  striking 
 analogies between quantities in such two states;
the vacuum $|0 \!\! >\leftrightarrow\Phi$, the pairing operator  
$ c_{k\uparrow}^{\dagger}c_{-k\downarrow}^{\dagger}\leftrightarrow \frac{\theta_i\theta_j}{z_i-z_j}$, and the coherence factor $g_k\leftrightarrow 1$.  
(The  factor $m$  on the SUSY Laughlin wavefunction is simply a scaling factor,
 and should  not be taken as the coherence factor. ) 
While the BCS vacuum  is  the  electron zero number  state, the corresponding ``vacuum'' of the SUSY 
QHE is  not  the $0$-particle state  but the original Laughlin state. The pairing operator in SUSY Laughlin wavefunction is interpreted as the 
 $p$-wave pairing  operator with adding spin-up degrees of freedom. 
Thus, the SUSY Laughlin wavefunction is regarded as an exotic   
$p$-wave pairing superfluid state on the vacuum Laughlin  state.  

The SUSY Laughlin wavefunction carries two important aspects, one of which is  QH aspect and the 
 other  is  BCS aspect.
When we define the fermion number operator  $F=\sum_i(\theta_i\frac{\partial}{\partial \theta_i}
-\theta_i^*\frac{\partial}{\partial \theta_i^*})$, 
the original Laughlin state is the vacuum  of  the fermion  $F\Phi=0$, and is the minimum fermion number state.
Meanwhile, our Moore-Read state $\Psi_{N/2}$ has the maximum fermion number as $F\Psi_{N/2}=N\Psi_{N/2}$.
Then,  the bosonic limit  corresponds to  the original Laughlin state, which reflects the original QH aspect, 
while the fermionic limit corresponds to the 
Moore-Read state, which reflects  BCS fermion pairing aspect. 
The coherence factor $g_k$ in the BCS wavefunction  represents the ratio of the  
amplitudes between the 0-particle state  and  the occupied state.  
When the coherence factor takes  unity,  the BCS state exists at the ``middle'' between the 0-particle and the occupied states, and the particle number fluctuation  is  maximized. As seen  above,    the coherence factor in the SUSY QHE reads  as  unity, 
which means the SUSY Laughlin wavefunction  has same contributions from  bosonic and fermionic sides.
This indicates that the SUSY Laughlin wavefunction possesses the supersymmetry.
Thus,   
 the existence of the SUSY in the present QH system could be physically ``translated''   in the language of the superfluidity. 

Though the Laughlin and the Moore-Read states belong to different topological order classes, they 
 are related as  component wavefunctions in a single SUSY Laughlin wavefunction.
The SUSY extension seems to circumvent the conventional no-go topological order arguments. 
The connections between the supersymmetrization and the topological order should be 
 pursued in a future research.

 Finally, we point out interesting analogies between the SUSY QHE and the supertwistor theory.
In both twistor and spherical QHE contain the Hopf map as a crucial ingredience of 
 thier constructions \cite{cond-mat/0211679}.
Further,  
the Grassmann coordinates in the supertwistor theory are also interpreted as spin
 degrees of freedom, and the 
  component wavefunctions of the 
 SUSY Laughlin wavefunction correspond to the twistor functions in the supertwistor formalism \cite{Prog.Theor.Phys.70181983}.
It would be worthwhile to speculate implications of  
 the correpondences between these two theories.

\vspace{2mm}
I would like to thank Masatoshi Sato and Shoucheng Zhang for valuable suggestions and 
 telling me  Ref.\cite{Prog.Theor.Phys.70181983}. 
This work was partially supported by the visiting program of ISSP in 2006.

\end{document}